\documentclass[superscriptaddress,amsmath,amssymb,twocolumn,showpacs,reprint,aps,floatfix,prl]{revtex4-1}
\usepackage{graphicx}
\usepackage{multirow}
\usepackage{dcolumn}
\usepackage{bm}

\newcounter{No}

\begin{document}

\title{Fission properties of $^{253}$Rf and the stability of neutron-deficient Rf isotopes}

\author{ A. Lopez-Martens}
\email{corresponding author: araceli.lopez-martens@ijclab.in2p3.fr}
\affiliation{IJClab, Universit\'e Paris Saclay and CNRS, F-91405 Orsay, France}
 \author{ K. Hauschild}
 \affiliation{IJClab, Universit\'e Paris Saclay and CNRS, F-91405 Orsay, France}
\author{ A.I. Svirikhin}
\affiliation{FLNR, JINR, Dubna, Russia}
\affiliation{Dubna State University, Dubna, Russia}

\author{ Z. Asfari }
 \affiliation{Universit\'e de Strasbourg, CNRS, IPHC UMR 7178, F-67000 Strasbourg}
   \author{M.L. Chelnokov}
 \affiliation{FLNR, JINR, Dubna, Russia}
 \author{V.I. Chepigin}
 \affiliation{FLNR, JINR, Dubna, Russia}
 \author{O. Dorvaux}
 \affiliation{Universit\'e de Strasbourg, CNRS, IPHC UMR 7178, F-67000 Strasbourg}
  \author{M. Forge}
  \affiliation{Universit\'e de Strasbourg, CNRS, IPHC UMR 7178, F-67000 Strasbourg}
\author{ B. Gall}
 \affiliation{Universit\'e de Strasbourg, CNRS, IPHC UMR 7178, F-67000 Strasbourg}

   \author{A.V. Isaev}
 \affiliation{FLNR, JINR, Dubna, Russia}
  \author{I.N. Izosimov}
 \affiliation{FLNR, JINR, Dubna, Russia}
 \author{K. Kessaci}
  \affiliation{Universit\'e de Strasbourg, CNRS, IPHC UMR 7178, F-67000 Strasbourg}
 \author{ A.A. Kuznetsova}
 \affiliation{FLNR, JINR, Dubna, Russia}
 \author{O.N. Malyshev}
 \affiliation{FLNR, JINR, Dubna, Russia}
  \author{R.S Mukhin}
 \affiliation{FLNR, JINR, Dubna, Russia}
 \author{ A.G. Popeko}
 \affiliation{FLNR, JINR, Dubna, Russia}
 \affiliation{Dubna State University, Dubna, Russia}

 \author{Yu.A. Popov}
  \affiliation{FLNR, JINR, Dubna, Russia}
  \author{B. Sailaubekov}
   \affiliation{FLNR, JINR, Dubna, Russia}
   \affiliation{L.N. Gumilyov Eurasian National University, 2 Satpayev Str, Nur-Sultan, Kazakhstan}
 \author{E.A. Sokol}
\affiliation{FLNR, JINR, Dubna, Russia}
\author{M.S. Tezekbayeva}
\affiliation{FLNR, JINR, Dubna, Russia}
\affiliation{The Institute of Nuclear Physics, 050032, Almaty, The Republic of Kazakhstan}
\author{ A.V. Yeremin} 
\affiliation{FLNR, JINR, Dubna, Russia}

\begin{abstract}
\begin{description}
\item[Background]  An analysis of recent experimental data [J. Khuyagbaatar et al., Phys. Rev. C {\bf104}, L031303 (2021)] has established the existence of two fissioning states in $^{253}$Rf: the ground state and a low-lying isomeric state, most likely involving the same neutron single-particle configurations as in the lighter isotone $^{251}$No. The ratio of fission half-lives measured in $^{253}$Rf was used to predict the fission properties of the 1/2$^{+}$ isomeric state in $^{251}$No and draw conclusions as to the stability against fission of even lighter Rf systems. 
\item[Purpose]     This paper focusses again on the fission properties of $^{253}$Rf and their impact on the stability of other neutron deficient isotopes, using new and improved data collected from two experiments performed at the Flerov Laboratory of Nuclear Reactions in Dubna, Russia. 
\item[Methods]    $^{253}$Rf and $^{251}$No nuclei were produced in fusion-evaporation reactions between $^{50}$Ti and $^{48}$Ca ions and the atoms of isotopically-enriched $^{204}$Pb targets. The nuclei of interest were separated from the background of other reaction products and implanted into a Si detector, where their characteristic radioactive decays were observed through position and time correlations between detected signals.
\item[Results]    Two fission activities with half-lives of 52.8(4.4)$~\mu$s and 9.9(1.2) ms were measured in the case of $^{253}$Rf, confirming the results of J. Kkuyagbaatar et al. A third state, at much higher excitation energy, was also observed through the detection of its electromagnetic decay to the 52.8$~\mu$s state. This observation leads to the opposite quantum-configuration assignments for the fissioning states as compared to the ones established by J. Khuyagbaatar et al., namely that the higher-spin state has the shortest fission half-life. This inversion of the ratio of fission hindrances between the low and high-spin states is corroborated in the isotone $^{251}$No by the non observation of any substantial fission branch from the low-spin isomer.
\item[Conclusions] In going from $^{251}$No to $^{253}$Rf, the fission half-life of a specific quantum state is found to decrease by close to 7 orders of magnitude.  Large reductions of more than 5 and 6 orders of magnitude are also found between the fission half-lives of the ground states of $^{252}$No and $^{254}$Rf and between those of $^{254}$No and $^{256}$Rf respectively, pointing to a similar rate of decrease of the fission barrier as one removes neutrons from both systems. Following this trend to the N=148 isotones, our results suggest that the fission half-life of the ground-state of the next even-even Rf isotope, $^{252}$Rf, will be extremely short, possibly at the limit of existence of an atom.

\end{description}
\end{abstract}

\
\maketitle

One of the main goals of modern nuclear-structure physics is to determine the limits of nuclear stability. For light nuclei, the limits of stability are determined by the neutron and proton drip lines, beyond which atomic nuclei decay by particle emission. For the heaviest systems, nuclear stability is governed by the competition between the surface tension of the nuclear liquid drop, which tends to hold the system together, and the strong Coulomb repulsion between the numerous protons, which drives the nucleus towards spontaneous fission. Increased stability against spontaneous fission arises because of quantum shell effects, which depend on the local density of single-particle states. High densities give rise to positive shell correction, meaning less stability, and low densities translate into enhanced stability. Alternating stabilising effects may coexist on the pathway to scission as a function of nuclear deformation. For heavy nuclei beyond Pu, negative shell effects bind the ground state by more than several MeV with respect to the liquid drop energy, while positive shell effects increase the energy of the system at the saddle point \cite{Bjornholm}. In $^{254}$No, the net result is a potential-energy barrier against fission of the order of 6.6 MeV while the liquid drop value is $\approx$0.9 MeV \cite{Henning}.\\
Not only the height but also the width and the structure of the barrier in multi-dimensional deformation space determine the fission half-lives. Other effects also come into play, such as the superfluidity or stiffness of the system in the fission process and the conservation of quantum numbers at level crossings on the path to fission (specialization energy). These effects are clearly demonstrated in the much longer spontaneous fission half-lives measured in nuclei with an odd number of nucleons as compared to the neighbouring even systems (see fig. 19 of ref. \cite{Hessberger17}, which shows fission hindrances typically varying between 10$^3$ and 10$^6$). When available, theoretical specialization energies in deformed actinides are found to strongly vary with the quantum numbers of the odd nucleonic configuration as well as its distance to the Fermi surface as the nuclear system evolves towards fission (see Ref. \cite{Koh} and references therein).  $^{253}$Rf offers therefore a unique possibility to study the effect of the occupation of specific neutron orbitals on fission times as the existence of two fissioning states, lying very close in excitation energy, has recently been confirmed in Ref. \cite{Baatar}. 

In this letter, new data on the decay properties of states in $^{253}$Rf are presented. The Nilsson configurations of the fissioning states of $^{253}$Rf in Ref. \cite{Baatar} are revised and the consequences in terms of the fission properties of states in $^{251}$No and $^{252}$Rf are discussed. 


The nucleus $^{253}$Rf was produced in the reaction $^{204}$Pb($^{50}$Ti,1n). The $^{50}$Ti beam was provided by the U400 Cyclotron at the Flerov Laboratory of Nuclear Reactions in Dubna using the Metal Ions from Volatile Compounds (MIVOC) method \cite{Rubert}. The average energy of the beam was 244 MeV.
The separator for heavy element spectroscopy (SHELS) \cite{SHELS} was used to select the evaporation residues of interest and the newly upgraded GABRIELA detector array \cite{Hauschild06,Chakma} was used to detect the residues and their subsequent $\alpha$ decay or fission as well as $\gamma$ and X rays and conversion electrons. After selection in SHELS, the evaporation residues of interest pass through a Time of Flight (ToF) detector and are then implanted into a 10$\times$10~cm$^{2}$ 500 $\mu$m thick Double-sided Silicon Strip Detector (DSSD). With 128 strips on both sides, this DSSD provides 16384 individual pixels for position and time correlation of the implanted ion with subsequent decays. Particles escaping from the DSSD are detected in eight 6$\times$5~cm$^{2}$ DSSDs with 16 strips on each side positioned upstream from the implantation detector and forming a tunnel. A Mylar degrader foil positioned at the entrance of the Si-detector array provides additional suppression of low-energy reaction products and insures shallow implantation of evaporation residues in order to perform conversion electron spectroscopy in the tunnel detectors. The Ge array consists of a large clover detector installed just behind the DSSD and 4 coaxial detectors forming a cross around the DSSD. All Ge detectors are equipped with new specially-designed BGO shields. Enriched $^{204}$PbS targets of 0.5 mg/cm$^2$ thickness on a 1.5 $\mu$m Ti backing were used. The enrichment of the target material was 99.94$\%$ (with 0.04$\%$ $^{206}$Pb, 0.01$\%$ $^{207}$Pb and 0.01$\%$ $^{208}$Pb impurities).  Calibrations of the DSSD were performed by implanting $^{209,210}$Ra and $^{216,217}$Th isotopes produced with  $^{164}$Dy and $^{170}$Er targets respectively and observing their characteristic $\alpha$ decay. The energy resolution of the DSSD is 15-20 keV for $\alpha$-particles ranging from 6-10 MeV. The calibration of the Ge detectors was performed using standard sources such as $^{152}$Eu and $^{133}$Ba, while the tunnel detectors were calibrated using the conversion electrons emitted in the decay of the 117~$\mu$s isomer in $^{209}$Ra \cite{Hauschild08}. 


\begin{figure}[htbp]
  \centering
   \includegraphics [angle=0,width=0.4\textwidth]{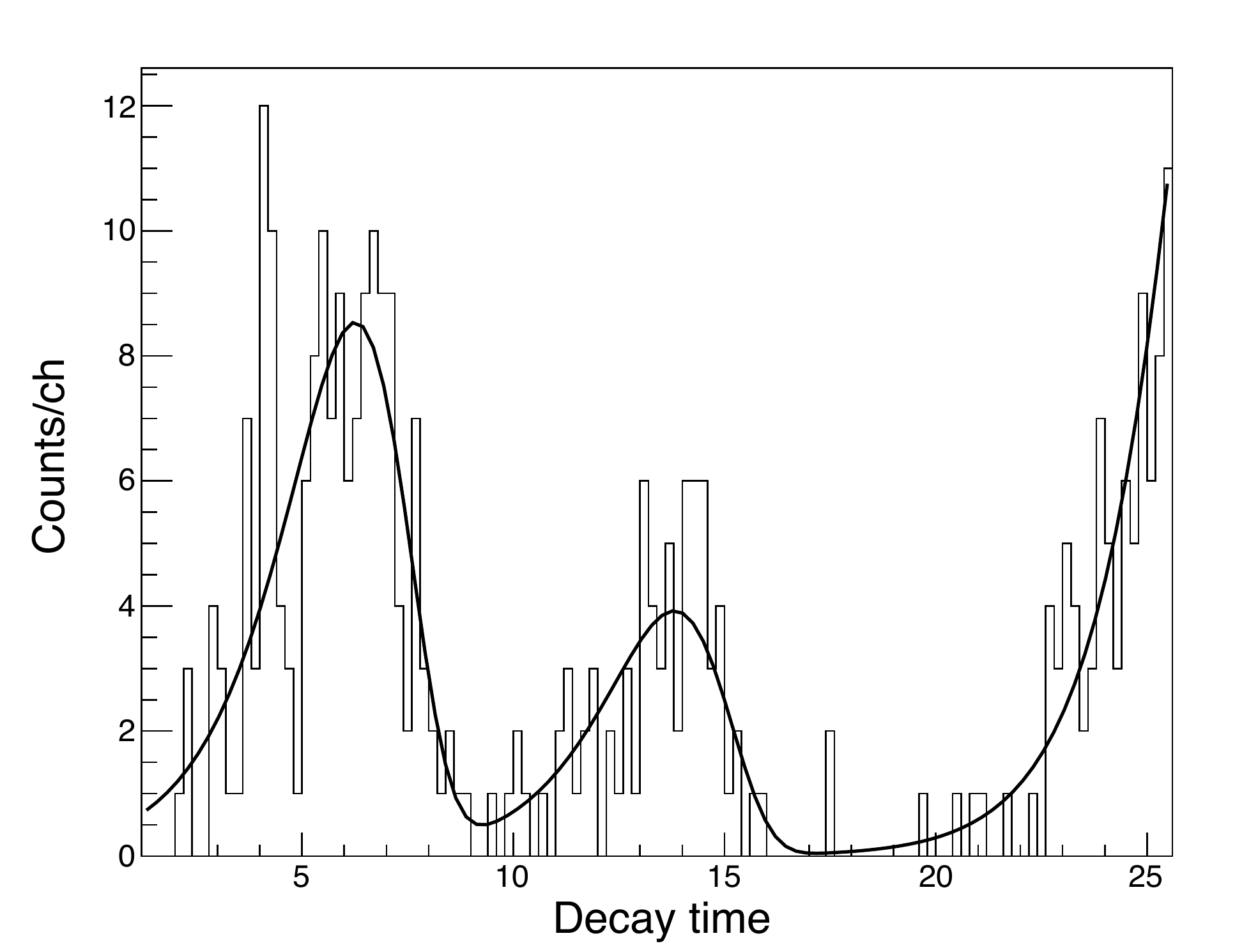} 
   \caption{Experimental distribution of fission times (in log$_2$($\mu$s) units) recorded during the experiment.  The black curve is a 3-component radioactive-decay fit to the distribution that accounts for the 2 fission activities as well as the tail of the random correlations.
   }
   \label{fig1}
\end{figure}
A total beam dose of 3$\times$10$^{18}$ particles was delivered onto the target during the course of the experiment. Fig. \ref{fig1} shows the time distribution of the 240 fission events observed. As in Ref. \cite{Baatar}, two groups of fission activities are clearly observed well below the random correlations, which set in above log$_2$(Time)$\approx$20 (or $\approx$1$~$s). The half-lives of the fission activities were measured to be 52.8(4.4)$~\mu$s and 9.9(1.2) ms, in agreement with the results of Ref. \cite{Baatar}. 
\begin{figure}[htbp]
  \centering
   \includegraphics [angle=0,width=0.5\textwidth]{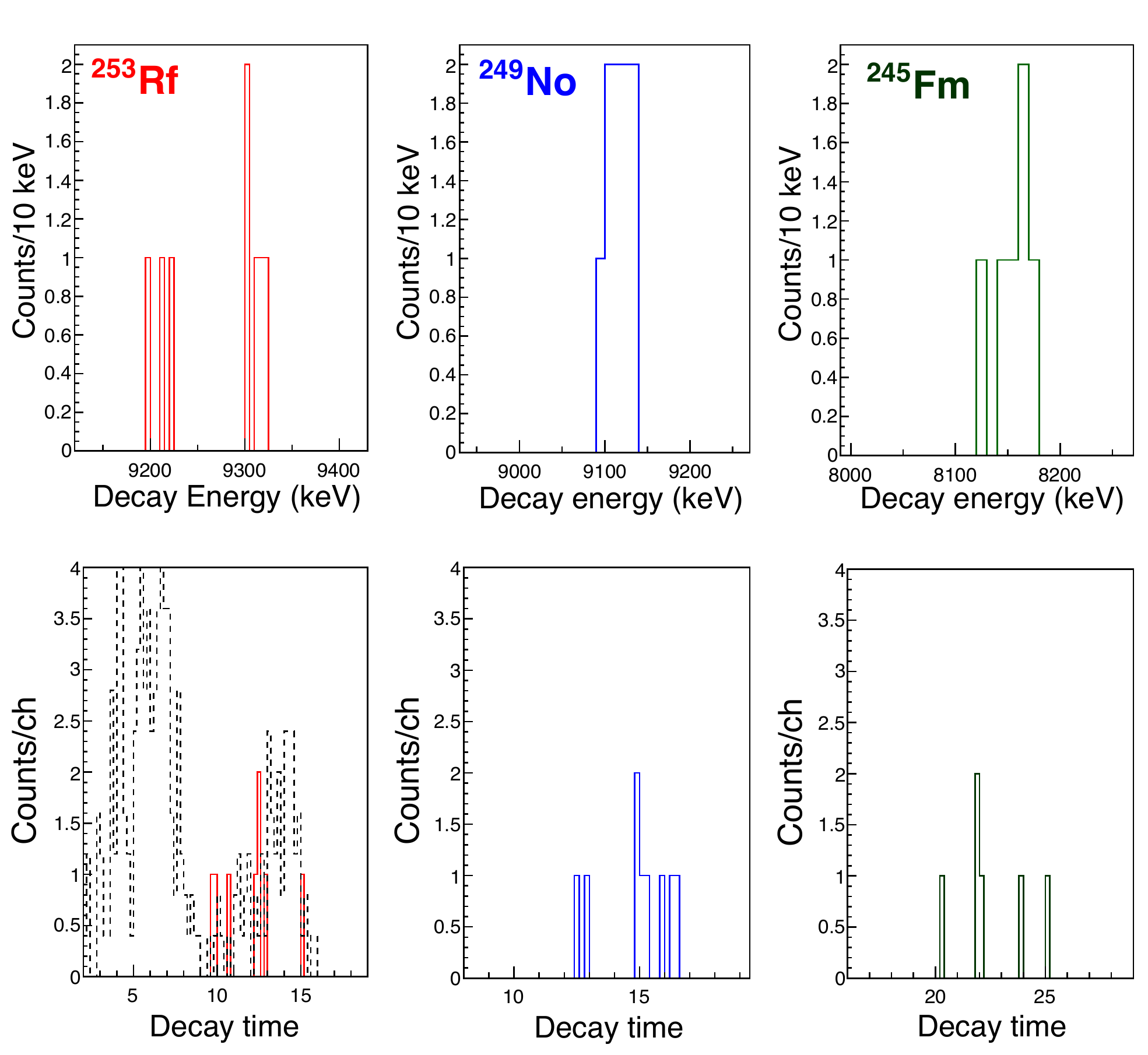} 
   \caption{Top row from left to right: High-energy part of the spectrum of alpha-particle energies obtained for the first, second and third generation decays following the implantation of an evaporation residue and detected in the same pixel of the implantation DSSD. The spectra were constructed with the unique condition that the second-generation decay has a decay time less than 300 ms. No energy conditions were used to search for $\alpha$-decay correlations. Bottom row from left to right: Time distribution (in log$_2$($\mu$s) units) of the high-energy alpha particles detected in the first, second and third generation decays. In the case of the first generation decays, the fission time distribution of Fig. \ref{fig1} is also displayed in dashed line.
   }
   \label{fig2}
\end{figure}
Fig. \ref{fig2} shows the observed position-correlated full-energy alpha decays, conditioned by the detection of at least two subsequent alpha decays, the first of which is imposed to have a half-life less than 300 ms. The figure also shows the corresponding distribution of decay times of the full-energy events of all three generations. In all cases, the random correlations occur on much longer time scales. The half-lives extracted using the procedure of Ref. \cite{KHS} are found to be: 5.7$^{+3.1}_{-1.5}~$ms for the first generation decays, 28.7$^{+14.3}_{-7.2}~$ms and 7.7$^{+5.3}_{-2.3}~$s for the  second and third generation decays respectively. The alpha activity of $^{253}$Rf is clearly revealed, since it is found to be followed by the characteristic $\approx$30 ms 9.13 MeV alpha decay of the recently discovered $^{249}$No \cite{Pepan} and the long-lived 8.17 MeV alpha decay of $^{245}$Fm \cite{Pepan}.  As can be deduced from the extracted half-life and as is illustrated in the bottom left panel of Fig. \ref {fig2}, the time distribution of the alpha activity of $^{253}$Rf is found to coincide with the time distribution of the slow fission activity of Fig. \ref{fig1}. From the ratio of alpha-decay and fission events, an alpha branch of B$_{\alpha}$=17(6)$\%$ can be deduced for the slow-fissioning state of $^{253}$Rf. The alpha decay energies of $^{253}$Rf appear in two different groups at 9.21(2) and 9.31(2) MeV and indicate that the decay populates more than one state in the daughter nucleus $^{249}$No, which is at variance with the decay scheme established in Ref. \cite{Baatar}.
\begin{figure}[htbp]
  \centering
   \includegraphics [angle=0,width=0.4\textwidth]{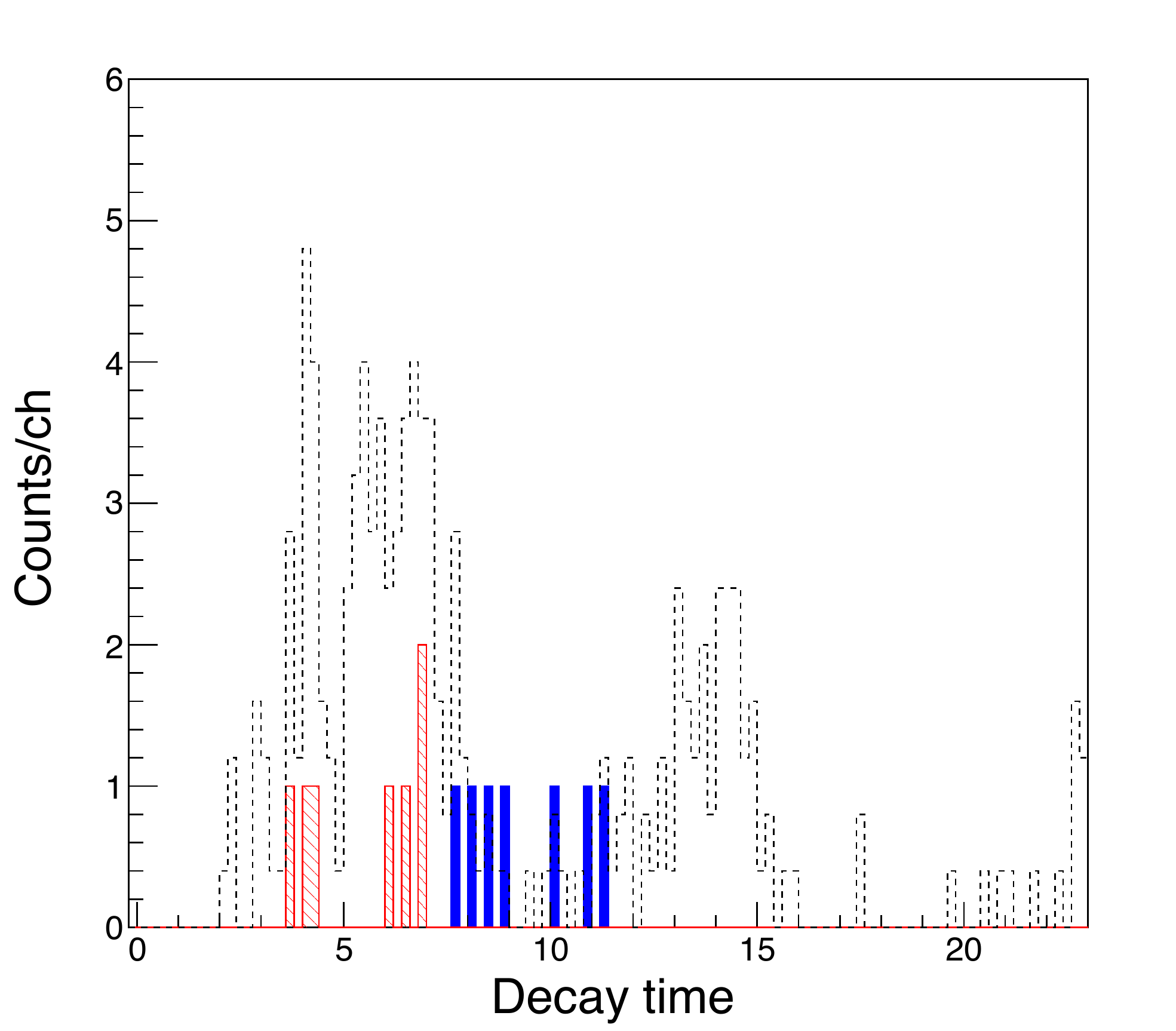} 
   \caption{Filled histogram: Time distribution (in log$_2$($\mu$s) units) of isomeric events. Dashed histogram: Time distribution of fission events following the isomeric decay. Dotted histogram: Time distribution of fission events of Fig. \ref{fig1}.
   }
   \label{fig3}
\end{figure}
Finally, as shown in Fig. \ref{fig3}, seven isomeric decays followed by the fast fission of $^{253}$Rf were observed. Given the number of observed $^{253}$Rf radioactive decays and assuming a transmission efficiency through SHELS of 40$\%$, a total cross section of 0.14(4)$~$nb could be deduced for the production of $^{253}$Rf in the reaction $^{204}$Pb($^{50}$Ti,1n). Within error, this value is consistent with what was measured in Ref. \cite{Baatar}. The half-life of the isomeric state was measured to be 0.66$^{+40}_{-18}~$ms and the energy removed by the isomeric cascades (obtained by summing all the signals observed in the GABRIELA detectors) was found to extend up to 1.02 MeV. Given the available states at low and medium excitation energy in N=149 isotones, in order for such a high-lying state to be isomeric, it must be a high-K isomer \cite{Dracoulis}.

\begin{figure}[t]
  \centering
   \includegraphics [angle=0,width=0.45\textwidth]{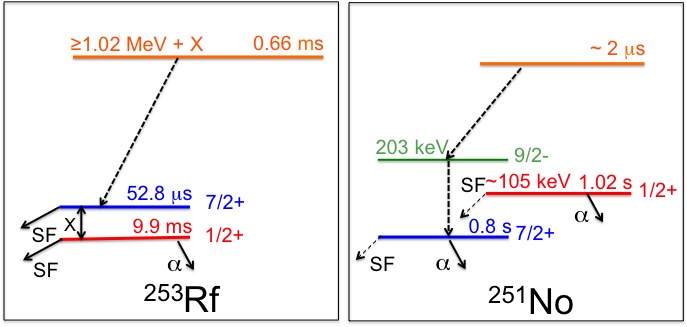} 
   \caption{Partial decay scheme established in this work for $^{253}$Rf and taken from \cite{Hessberger06} for $^{251}$No. See text for details. }
   \label{fig4}
\end{figure}

Based on the systematics of N=149 isotones, the ground state and first excited state of $^{253}$Rf are expected to be based on either the 7/2$^+$[724] or 1/2$^+$[631] Nilsson configurations. The fact that the 0.66$~$ms isomer decay is found to be followed only by fast fission events indicates that the fast fissioning state must have higher spin than the slow-fissioning state. Such a situation is also observed in $^{251}$No (see Fig. \ref{fig4}), where the $\approx$2$~\mu$s isomer is observed to decay to the 9/2$^-$ band head and then to the 7/2$^{+}$ ground state, bypassing the 1/2$^{+}$ isomer \cite{Hessberger06}. Therefore, and contrary to what is concluded in Ref. \cite{Baatar}, the 52.8$~\mu$s state in $^{253}$Rf is assigned to the 7/2$^+$[724] configuration and the state with the longer spontaneous fission half-life and 17(6)$\%$ alpha-decay branch is assigned to the 1/2$^+$[631] configuration. As discussed in Ref. \cite{Baatar}, the relative positions of the 1/2$^+$ and 7/2$^+$ states in $^{253}$Rf are uncertain, but if the 1/2$^+$ is the ground-state, the 7/2$^+$ state must lie at an excitation energy lower or less than a few keV higher than the 3/2$^+$ member of the ground-state band in order to have the measured half-life. Given the decoupling parameter of the 1/2$^+$[631] configuration known from lighter isotopes, this in turn sets the difference in energy between the two states X$\lesssim$15 keV (see Fig. \ref{fig4}), with the possibility of X being negative and the 7/2$^+$ state being in fact the ground state. Moreover, the down-sloping trend of the 1/2$^+$ state in N=149 isotones as Z increases sets a probable lower limit of $\approx$-105$~$keV for X.

\begin{figure}[htbp]
  \centering
   \includegraphics [angle=0,width=0.5\textwidth]{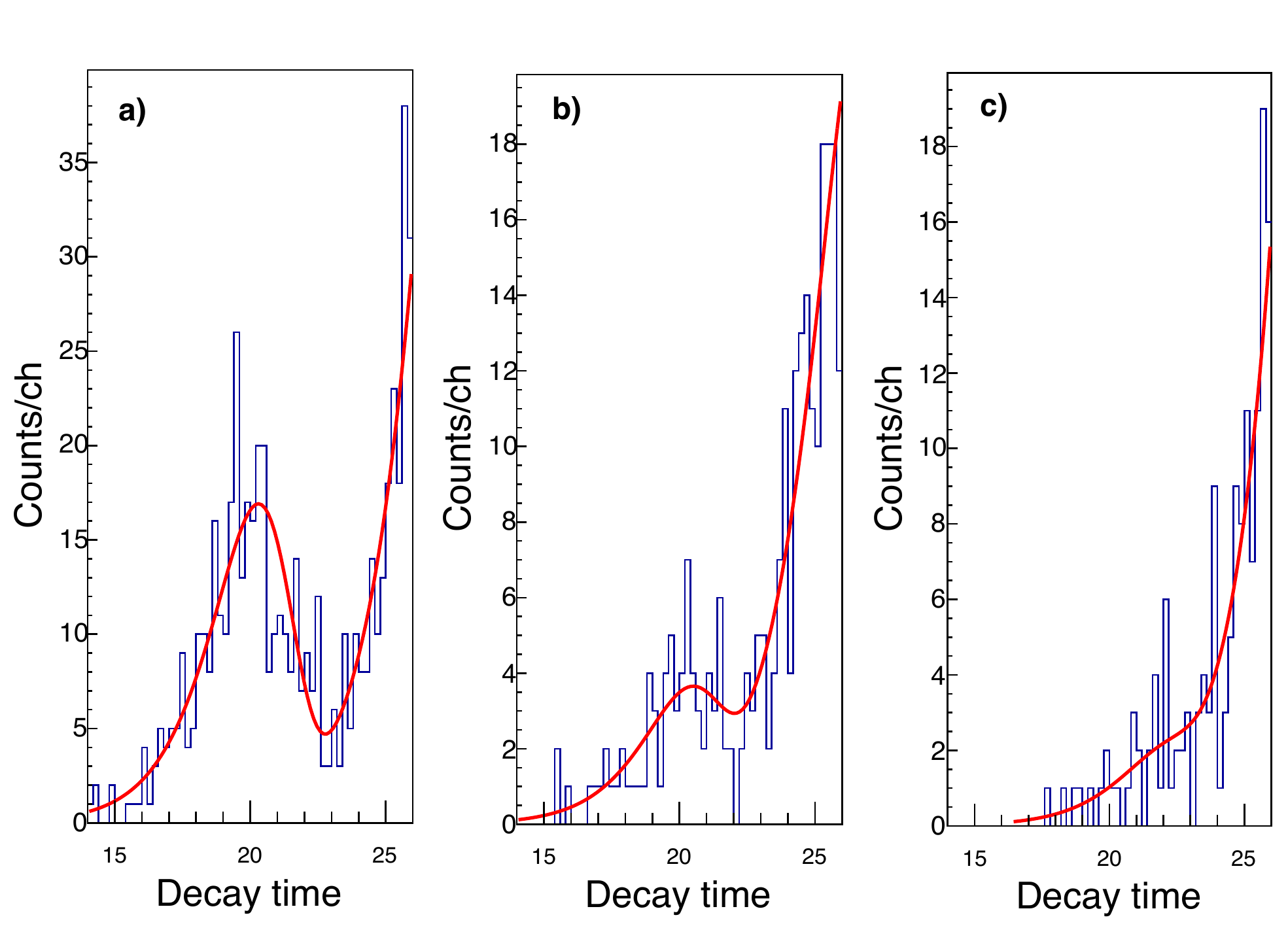} 
   \caption{a) Time distribution (in log$_2$($\mu$s) units) of ground-state $^{251}$No alpha decays, of b) isomeric $^{251m}$No alpha decays and of c) all the fission events detected in the reaction $^{204}$Pb($^{48}$Ca,xn)$^{252-x}$No \cite{Pepan}.}
   \label{fig5}
\end{figure}

Given the different spin assignments of the two aforementioned states as compared to what was suggested in Ref. \cite{Baatar}, the difference in fission hindrance  $\Delta F_H$=T$_{SF,1/2^+}$/T$_{SF,7/2^+}$= 226(33) has a different consequence on the estimate of the spontaneous fission half-life of the 1/2$^+$ isomer in $^{251}$No. Indeed, instead of the 1.7$~$s spontaneous fission half-life estimated in Ref. \cite{Baatar}, we find $T_{SF}(^{251m}No)$ $\approx$36 h. This is supported by our data on $^{249-251}$No, in which the $^{249}$No isotope was first discovered \cite{Pepan}. In the experiment described in Ref. \cite{Pepan}, nobelium nuclei were produced in complete fusion reactions of a $^{48}$Ca beam and the same $^{204}$PbS target as used for the $^{253}$Rf production. Fig. \ref{fig5} shows the time distribution of ground-state and isomeric alpha decays of $^{251}$No together with the time distribution of all the fission events observed in that experiment. The two-component fit of the fission time distribution in panel c) yields 
23(7) fission events with a half-life of 2.2(8)$~$s on top of the tail of the random correlations. This number of fission events and associated half-life is perfectly in line with the measured production of $^{252}$No nuclei on $^{206}$Pb impurities of the target, which yielded a total of ten $^{252}$No$-^{248}$Fm mother-daughter alpha-decay correlations, as well as with the 318(18) $^{251}$No ground-state alpha decays extracted from the fit of Fig. \ref{fig5}a). Indeed, given the known fission branches of the ground states of $^{251}$No and $^{252}$No \cite{NNDC,Hessberger06}, we expect of the order of twenty two $^{252}$No fission events and two $^{251}$No ground-state fission events.  There is therefore no room for any substantial fission branch of $^{251m}$No, contrary to what is  claimed in Ref. \cite{Baatar}. 
Regarding the conclusion of Ref. \cite{Baatar} that there is no abrupt fall of fission half-lives in odd-A No and Rf isotopes, this work shows, on the contrary, that  in going from $^{251}$No to $^{253}$Rf, i.e by adding two protons to the system, the 7/2$^+$ state, which is the ground-state in $^{251}$No and either a very low-lying excited state or even still the ground-state in $^{253}$Rf (see Fig. \ref{fig4}), undergoes a dramatic change in fission properties. In $^{251}$No, it has a spontaneous fission half-life of 571$~$s and in $^{253}$Rf its spontaneous fission half-life is 52.8$~\mu$s. This represents the largest known reduction in spontaneous fission half-life between identical states in neutron-deficient No and Rf isotones, as shown in Table \ref{tab1}. For N=148 isotones, the fission half-life is only known for $^{250}$No. Applying the average reduction factor of 5.3$\times$10$^{6}$ between No and Rf fission half-lives in Tab. \ref{tab1}, an estimate of 6.9$\times$10$^{-13}~$s is obtained for the spontaneous fission half-life of the ground state of $^{252}$Rf.
\begin{table}[htbp]
\caption{Comparison of spontaneous fissions half-lives of nuclear states in No and Rf isotopes. Level half-lives, energies and spins are taken from Ref. \cite{Hessberger06} and this work.}
\begin{center}
\begin{tabular}{|c c c c c|}
\hline
Z & N &  E$^*$ & I$^{\pi}$ & T$_{SF}$ \\
\hline\hline
 
102 & 152 &  0 & 0$^+$ & 3$\times$10$^{4}$  s\\
104 & 152  & 0 & 0$^+$ & 6.4$\times$10$^{-3}$  s\\
\hline
102 & 150 & 0 & 0$^+$ & 8.43 s  \\
104 & 150  & 0 & 0$^+$ & 23$\times$10$^{-6}$  s\\
\hline
102 & 149 & 0 & 7/2$^+$ & 571 s \\
104 & 149 & $<$15 keV & 7/2$^+$ & 52.8$\times$10$^{-6}$  s\\
\hline
102 & 148 & 0 & 0$^+$ & 3.7$\times$10$^{-6}$ s \\
104 & 148 & 0 & 0$^+$ & not known \\
\hline

\end{tabular}
\end{center}
\label{tab1}
\end{table}%


In this work, we have confirmed the fission properties of two states in $^{253}$Rf. The spontaneous fission half-lives of these two states, which lie within less than 100$~$keV of each, differ by a factor $\approx$220. This is a manifestation of different specialization energies, making the fission barrier of one state higher and hence increasing its fission time with respect to the other. From the decay properties of a higher-lying isomeric state, we have determined that the shorter-lived fission activity corresponds, in fact, to the 7/2$^+$[624] neutron configuration, and the longer-lived activity to the 1/2$^+$[631] configuration. This in turn leads to radically different conclusions to the ones exposed in Ref. \cite{Baatar}, namely that the 1/2$^+$ isomeric state in $^{251}$No cannot have a substantial fission branch and that there is in fact a fall in fission half-lives in neutron deficient Rf isotopes since the 7/2$^+$ state in $^{253}$Rf has a $\approx$10$^7$ shorter spontaneous fission half-life than in $^{251}$No. This is in line with the observed decrease in fission half-lives of heavier Rf isotopes (with neutron number N$\le$152) with respect to their corresponding No isotone. If we apply the average trend of decrease of fission half-lives from $^{250}$No to $^{252}$Rf isotones, then one should expect the ground-state of $^{252}$Rf to have a sub-ps spontaneous fission half-life, making it experimentally inaccessible and very close to the 10$^{-14}~$s limit of existence of an atom.

\begin{acknowledgments}

This work was supported by the Russian Foundation for Basic Research (project no. 17-02-00867), the French national Research Agency (projects nos. ANR-06-BLAN-0034-01 and ANR-12-BS05-0013) and the IN2P3-JINR collaboration agreement no. 04-63. 
\end{acknowledgments}

\end{document}